# Observations of four-wave mixing in slow-light silicon photonic crystal waveguides

James F. McMillan[1], Mingbin Yu[2], Dim-Lee Kwong[2], and Chee Wei Wong[1]

[1] *Optical Nanostructures Laboratory, Center for Integrated Science and Engineering, Solid-State Science and Engineering, and Mechanical Engineering, Columbia University, New York, NY 10027*
[2] *Institute of Microelectronics, Singapore, Singapore 117685*
*Author e-mail address:* jfm2113@columbia.edu, cww2104@columbia.edu

**Abstract:** Four-wave mixing is observed in a silicon W1 photonic crystal waveguide. The dispersion dependence of the idler conversion efficiency is measured and shown to be enhanced at wavelengths exhibiting slow group velocities. A 12-dB increase in the conversion efficiency is observed. Concurrently, a decrease in the conversion bandwidth is observed due to the increase in group velocity dispersion in the slow-light regime. The experimentally observed conversion efficiencies agree with the numerically modeled results.

**1. Introduction**

The investigation of nonlinear optical processes in silicon-on-insulator (SOI) waveguides has reached a level a maturity that has allowed researchers to move beyond the proof-of-concept phase and into the practical application realm. In particular, the third-order nonlinear process of four-wave mixing (FWM) in silicon integrated waveguides has recently received particularly intense investigation due to the large third-order susceptibility ($\chi^{(3)}$) of silicon and a wavelength-tunable method of signal amplification, conversion, and regeneration is an extremely attractive capability in the silicon-on-insulator platform for integrated photonic circuits.

Initial experimental investigations into FWM in silicon waveguides [1,2] were successful due to the large pump powers and small modal areas of the waveguides used. The small effective modal area of these waveguides ensured large optical power densities within the waveguides resulting in large nonlinear optical responses. Subsequent investigations [3-5] demonstrated the ability to optimize not only the conversion efficiency but also the available bandwidth. This was accomplished by dispersion engineering the waveguide: tuning the waveguide modal dispersion to optimize phase matching at the wavelengths of interest. With the modal dispersion optimized, maximum conversion efficiencies of up to -9.6 dB and usable bandwidths greater than 150 nm have been demonstrated [3]. In addition to these optimizations, further enhancement of the FWM conversion efficiency has been demonstrated utilizing ring resonators [1,6,7], hybrid silicon-organic slot waveguide structures [8], and chalcogenide-based planar waveguides [9]. These promising proof-of-concept experiments have paved the way for the demonstration of practical applications of FWM in silicon. These applications include: 40 GB/s [10], and 160 GB/s [11] have been demonstrated to date, other applications include all optical signal regeneration [12], spectral phase conjugation [13], and waveform compression [14]. The utility of four-wave mixing in silicon waveguides has branched out beyond applications grounded in optical communications system and into a number of other areas. The realization that FWM could be utilized as an on-chip optical time lens [15] has led to a number of interesting new applications including temporal [16] and spectral magnification [17] and the demonstration of an ultrafast optical oscilloscope [18].

For the majority of applications of FWM in silicon waveguides, the length of the waveguide needed to achieve a large conversion efficiency is typically on the order of several centimeters, occupying a large footprint on the photonic integrated circuit. Photonic crystal waveguides with slow-light modes, however, offer an approach to significantly increase the light-matter interaction with significantly (100× or more) smaller footprints. The modes of a photonic crystal waveguide exhibit unique dispersion properties that have been demonstrated to exhibit large group indices [19-22]. The slow-light modes of the two-dimensional, hexagonal photonic crystal single line-defect waveguides has been demonstrated to enhance a number of nonlinear phenomenon including Raman scattering [23-25], self-phase modulation [26-30], third-harmonic generation [31], in the presence of nonlinear absorption such as two-photon (TPA), three-photon, and free carrier absorption (FCA) [32].

In the specific case of slow-light enhancement of FWM in photonic crystal waveguides, the conversion efficiency has been theoretically suggested to be enhanced due to the increase in the effective length of the waveguide [33,34]. However, due to the large group velocity dispersion (GVD) of these waveguides in the slow-light regime, the conversion efficiency bandwidth is typically reduced. An enhancement of conversion efficiency was demonstrated in photonic crystal waveguides [35,36]. In addition, chalcogenide-based photonic crystal waveguide has observed conversion efficiencies 13× higher than in silicon wire waveguides [37]. In this paper, we experimentally demonstrate the slow-light enhancement of the FWM conversion efficiency and bandwidth in the silicon W1 photonic crystal waveguides. In addition, we have mapped out the wavelength dependence of the enhancement and investigated the enhancements dependence on group index.

## 2. Theory

We consider partially degenerate four-wave mixing (FWM) in our experiment, where a pump wave with optical frequency $\omega_{pump}$ couples power with a signal wave with frequency $\omega_{signal}$ via the third-order nonlinear susceptibility ($\chi^{(3)}$) of silicon into an idler wave with an optical frequency determined by $\omega_{idler} = 2\omega_{pump} - \omega_{signal}$. The efficiency of this exchange of energy is dependent on the level of coherence between all the frequencies involved ($\omega_{pump}$, $\omega_{signal}$, $\omega_{idler}$) in order to maintain conservation of momentum. For optical waves confined to a guiding mode structure, such as a photonic crystal waveguide, the coherence is determined by the nonlinear and linear phase mismatch of the propagating modes. The lower the phase mismatch, the more efficient the conversion of power from to $\omega_{idler}$.

The goal of this experiment is to characterize the wavelength dependence of this conversion efficiency within a silicon W1 photonic crystal waveguide (PhCWG). Due to the large group velocity dispersion in these waveguides, it is expected that the efficiency of the power conversion between the signal wave and the idler wave, $G_{idler}$, will be highly wavelength dependent on the pump wavelength and the signal detuning from the pump. By extending previous work done in fibers [38], it has been shown that for a photonic crystal waveguide of length $L$, the conversion efficiency can be written as [33]:

$$G_{idler} = \frac{P_{idler}^{out}}{P_{signal}^{in}} = \left(\frac{\gamma^* \overline{P}_p}{g} \sinh(gL)\right)^2 e^{-\alpha_{idler}^* L} \quad (1)$$

The efficiency is governed by an effective coupled pump power, $\overline{P}_P = P(0)\left(1 - e^{-\alpha_{pump}^* L}\right) / \alpha_{pump}^* L$, which accounts for the losses experienced by the pump wave. In Eq. 1 the parametric gain, g, is the term which carries the phase dependent variables:

$$g = \sqrt{(\gamma \overline{P}_p)^2 - \left(\frac{\Delta k_L + \Delta k_{NL}}{2}\right)^2} \quad (2)$$

The linear phase mismatch between the interacting optical frequencies is $\Delta k_L = 2k_{pump} - k_{signal} - k_{idler}$. In order to examine the wavelength dependence of $\Delta k_L$ it can be approximated using the Taylor expansion about the pump frequency:

$$\Delta k_L \approx (\Delta w)^2 \beta_2(\lambda_{pump}) + \frac{1}{12}(\Delta w)^4 \beta_4(\lambda_{pump}) \quad (3)$$

where $\Delta\omega = |\omega_{pump} - \omega_{signal}|$. The dispersion dependent parameters, $\beta_2(\lambda)$ and $\beta_4(\lambda)$, can be derived [39] from the numerically computed [40] dispersion relation of the guided mode of interest within the PhCWG. In this work all wavelengths propagate in the fundamental quasi-TE guided mode of the PhCWG. The other phase term in Eq. 2, $\Delta k_{NL}$, is the nonlinear phase mismatch and is defined as

$$\Delta k_{NL} = 2\gamma^* \overline{P}_P \qquad (4)$$

From Eq. 1 and the dispersion parameters for the waveguide mode, we are able to confidently assess the validity of our experimentally measured data. One modification does need to be made in order to accurately model the effect of slow-light on the propagating mode. In the silicon PhCWG, as the wavelength approaches the modes onset wavelength ($\lambda_{os}$, e.g. the region of a low group velocity), the mode experiences strong localization, increasing the optical intensity within the material and at the same time the modes low group velocity gives it more of an opportunity to experience light-matter interaction like the $\chi^{(3)}$ process and out of plane scattering due to loss. To incorporate this phenomenon within our simple model, we substitute in Equations (1), (2), and (4), the nonlinear parameter ($\gamma$) and the waveguide propagation loss ($\alpha$) with group index dependent versions of themselves ($\gamma^*, \alpha^*$). For the third-order nonlinear parameter we used an $n_g^2$ dependence [26]: $\gamma^* = \gamma(n_g/n_0)^2$. The group index scaling of the propagation loss is a more subtle choice, since a number of various $n_g$ dependences have been suggested [41], including ones that depend on the group velocity regime [42,43]. We chose to use linear $n_g$ dependence in the model, so that $\alpha^* = \alpha_0(n_g/n_0)$, where a linear scaling of $\alpha$ was observed experimentally to fit well with our measurements, as detailed below. This simple model neglects nonlinear absorption effects however due to the relatively short length of our waveguide and the low coupled powers used, two-photon and free-carrier absorption effects are minimal in these particular measurements and the model provides basic trends of the underlying FWM.

## 3. Experiment

The W1 waveguide utilized in this experiment were fabricated using deep-UV photolithography on a silicon-on-insulator wafer with a silicon thickness of 250 nm and a buried oxide thickness of 1 um. The lattice constant (*a*) for the hexagonal photonic crystal membrane was 440 nm and the radius of the fabricated holes were 0.36a (~158 nm). The total length of the photonic crystal waveguide was 1000*a* (0.44 mm). The photonic crystal region was subjected to a hydrofluoric acid etch in order to remove the oxide beneath it, creating a suspended silicon membrane with air cladding above and below. The input and output of the PhCWG was coupled to channel waveguides 450 nm in width. The total length of the device, from input facet to output facet, was approximately 3 mm. The transmission of TE polarized light through the waveguide can be seen in seen in Fig. 1(a). The significant drop in transmission, which corresponds to the fundamental quasi-TE mode on-set wavelength ($\lambda_{os}$), occurs at roughly 1545.5 nm.

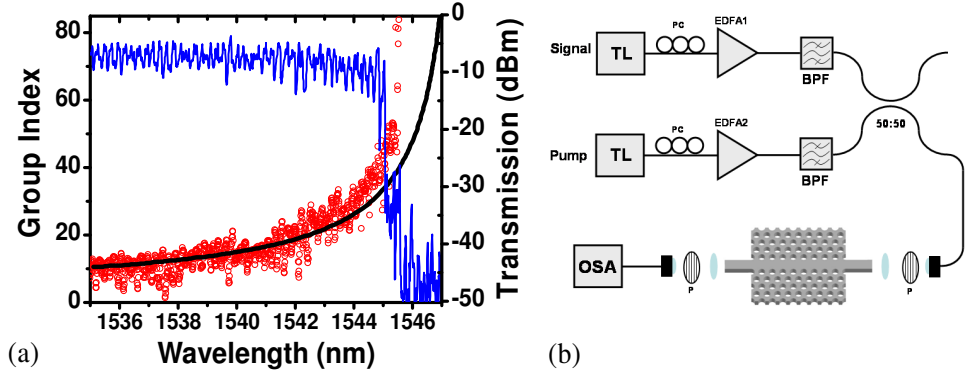

(a)                                                  (b)

Fig. 1. (a) The optical transmission (blue line), measured group velocity (red points), and fitted numerically calculated group velocity (black line). (b) Experimental setup for the measurement of partially degenerate four-wave mixing.

In order to determine the group index of the waveguide with respect to wavelength, we employed the phase-delay method [44] in which the light from a tunable laser was modulated at a frequency of 5 GHz using a lithium niobate modulator and microwave synthesizer. The light was then transmitted through the waveguide, collected, and fed into a high-speed photodetector (bandwidth~15 GHz). The wavelength dependent time delay ($\tau(\lambda)$) between the transmitted waveform and the modulator driving signal was measured using a high speed sampling oscilloscope. The measured delay was used to compute the group index by from the relation $n_g(\lambda) = n_{wg} + \Delta n(\lambda) = n_{wg} + c\Delta\tau(\lambda)/L_{PhCWG}$, where $\Delta\tau(\lambda)$ is the measured time delay difference between the PhCWG and a channel waveguide of the same length with a known group index ($n_{wg}$). The measured group index is seen to rapidly increase as the wavelength approaches $\lambda_{os}$=1545.5 nm, confirmation that this is indeed the onset of the quasi-TE mode, or more accurately this is the wavelength that is as close to the $\lambda_{os}$ that disorder will allow us to get due to disorder induced broadening of the mode onset.

Degenerate four-wave mixing in the PhCWG was observed using the experimental setup in Fig. 1(b). Two tunable laser sources, one acting as a pump and the other as a signal probe, are both amplified using separate erbium doped fiber amplifiers and then subsequently optically filtered (3-dB bandwidth ~ 0.2 nm) of amplified spontaneous emission noise. The pump and signal were combined using a 50:50 fiber coupler and the combined wavelengths were coupled into the waveguide using aspheric lenses (NA=0.6). The light exiting the waveguide was collimated back into fiber and analyzed using an optical spectrum analyzer with 10 pm resolution. Taking into account filter loss, collimator loss and an 8.5 dB per facet coupling loss, the estimated coupled pump power was +14 dBm.

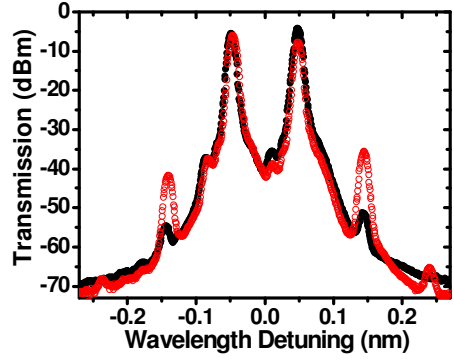

Fig. 2. Optical spectrum of degenerate four-wave mixing in a silicon photonic crystal waveguides. (black) $\lambda_{pump}$ = 1535.00 nm ($n_g \approx 10$), (red) $\lambda_{pump}$ = 1545.75 nm ($n_g \approx 80$). In both plots, the $\lambda_{signal}$ is blue shifted from $\lambda_{pump}$ by ~12.5 GHz.

Experimentally observed four-wave mixing in a photonic crystal waveguides for two different wavelengths ($\lambda_{pump}$ = 1535 nm and $\lambda_{pump}$ = 1545.75 nm) can be seen in Fig. 2. In both measured spectrums the difference between the pump and signal is roughly 12.5 GHz whereas the pump is the laser with the higher wavelength (the peak at +0.5 nm). As can be seen from these spectrums, the generated idler power for the case of the high group index (1545.75 nm, $n_g \sim 80$) is greater than that of lower group index (1535 nm, $n_g \sim 10$). As further evidence of increased FWM conversion efficiency, it can be seen in the high group index spectrum that the higher order idler (the small peaks at ±0.24 nm), which are generated from cascade parametric mixing of the signal and pump with their fundamental idler. Also observable in the high group index spectrum is what appears to be attenuation of the transmitted pump which is due in part to it experiencing a higher group index and also due to the strong Fabry-Perot oscillations at these wavelengths.

In order to investigate the wavelength dependence of the four wave mixing conversion efficiency the generated idler power at $2\omega_{pump} - \omega_{signal}$ was measured for signal wavelengths -5 to +5 nm detuned from the pump wavelength in 0.1 nm steps. This was done for pump wavelengths from 1535 to 1545.5 nm in 0.5 nm steps. The resulting two-dimensional map of measured idler powers can be seen in Fig. 3(a). In order to corroborate the measured results, the model outlined previously was calculated using the numerically computed dispersion, the result of which be seen in Fig. 3(b). Since experimentally we measure $P_{idler}^{out}$, in order to compare with our measurements we numerically computed $P_{signal}^{in} G_{idler}$ and accounted for experimental coupling losses. In order to be consistent, the model is computed with the same wavelength steps as the experimental data ($\Delta\lambda_{signal}$= 0.1 nm, $\Delta\lambda_{pump}$ = 0.5 nm). The free parameters used to fit the numerical model with the experimental data were the wavelength shift in the calculated group index and the minimum linear loss, $\alpha_0$. The best fit to experimental results were obtained with a 2.6% blue-shift of the calculated group index (the group index used in the model is plotted in Fig.1(a)) which could possibly be due to thinning of the silicon slab during the oxide etch [45], and an $\alpha_0$ of 10 dB/cm. The resulting value of $\alpha_0$ used in the model is lower than values we have measured using the cut-back method for similarly fabricated waveguides, but is not out of the realm of possibility for such a device. Slices from the 2D figures that directly compare the measured results and the numerical model for $\lambda_{pump}$ = 1535 nm and $\lambda_{pump}$ = 1545.5 nm can be seen in Fig. 3(c). The maximum measured idler power for each pump wavelength is plotted in Fig. 3(d).

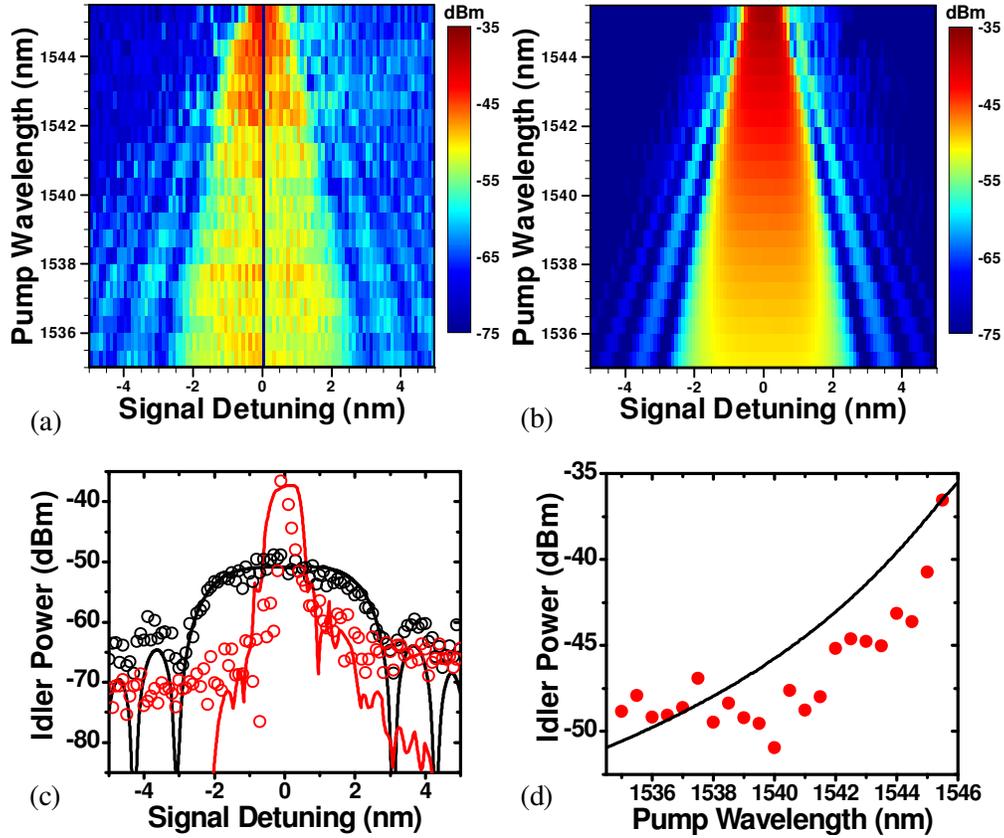

Fig. 3. (a) Measured idler power wavelength dependence ($\Delta\lambda_{pump}$ = 0.5 nm, $\Delta\lambda_{signal}$ = 0.1 nm) (b) Simulated idler power wavelength dependence (c) Idler power dependence for $\lambda_{pump}$ = 1535 nm (black) and $\lambda_{pump}$ = 1545.5 nm (red). (d) Maximum measured idler power for a given pump wavelength. (points: measurement, solid line: calculation)

The calculated wavelength dependence of the generated idler power matches closely with that of the measured results and the maximum generated idler power is observed to increase by over 12 dB, while the bandwidth over which a measurable idler signal is observed decreases as the pump group index increases. In addition, the measured and calculated results show a noticeable asymmetry as the pump wavelength approaches $\lambda_{os}$. This asymmetry can be attributed to the fact that for signal wavelengths blue shifted from the pump wavelength (signal detuning < 0 nm) generate idlers at higher group indices, which experience large propagation loss. This becomes more apparent as the pump wavelength approaches $\lambda_{os}$, as can be seen in the plot for $\lambda_{pump}$ = 1545.5 nm in Fig. 3(c) where the measured idler power is seen to drop by 14 dBm from +0.1 nm to -0.1 nm.

## 4. Results

In order to quantify the FWM process in the PhCWG, the measure idler powers should be normalized with respect to the signal power in order to gauge the conversion efficiency. The conversion efficiency is derived experimentally by taking the ratio of the measured output idler power and measured output signal power ($P^{out}_{idler} / P^{out}_{signal}$) in order to cancel any uncertainty in the coupling efficiency. It has recently been demonstrated [46] that using this definition can inadvertently overestimate conversion efficiency in silicon channel waveguides due to loss experienced by the signal within the waveguide, either through conventional linear propagation loss or by nonlinear loss due to two-photon and free-carrier absorption. In order to avoid this, the conversion efficiency can be defined as $P_{idler}^{out}/P_{signal}^{in}$. Using this definition, the pump wavelength dependence conversion efficiency of our waveguide can be derived from the maximum measured idler powers (See Fig. 3(d)) by accounting for signal input power and the idler coupling loss. Doing so gives a maximum measured conversion efficiency of -38.8 dB. We can conclude that this value is the lower bound estimate of the maximum conversion efficiency of our waveguide for this coupled pump power. The assumption made with this definition is that the coupling to the waveguide of both the pump and signal is wavelength independent, which is valid for channel waveguides for the bandwidths considered here but not for a photonic crystal waveguide which exhibits increased coupling losses as the mode onset wavelength is approached.

With the lower bound conversion efficiency of the experiment stated, we will now present the results for the conventional definition ($P_{idler}^{out}/P_{signal}^{out}$), with a caveat, in order to make an attempt at correcting for this group index dependent coupling loss. The stated caveat is that the conversion efficiencies will be derived from the maximum idler power generated by a signal detuned ±0.1 nm from the pump wavelength in order to minimize any difference in the group index dependent coupling loss experienced by the idler and signal. This is contrasted by the previous definition where the measured maximum measured idler power for a given pump wavelength, which due to the strong Fabry-Perot oscillations (~4 dB) in the waveguide, may not have been in such close proximity to the pump wavelength for the measured data.

The pump wavelength dependence of the conversion efficiency can be seen in Fig. 4(a). The maximum measured conversion efficiency, using $P_{idler}^{out}/P_{signal}^{out}$ definition, is -35.7 dB, a difference of ~3 dB compared with the lower bound. This 3dB difference stems from a combination of the wavelength dependent pump propagation and coupling losses which would lower the first definition of maximum conversion efficiency and also the increase in the signal propagation loss, which results in an artificial enhancement in the second definition. The numerical model appears to underestimate the conversion efficiency and the maximum idler power for lower wavelengths. It is believed this is due to a small amount of FWM occurring in the channel waveguide that couples to the PhCWG. This manifests in the experiment as a small amount of constant idler power that is wavelength independent due the fact that waveguides of these dimensions and length have low but relatively broadband, when compared to the PhCWG, conversion efficiencies [1]. In order to explicitly display the group index enhancement of the conversion efficiency it is plotted versus group index in Fig. 4(b). In plotting the experimentally measured conversion efficiency, the experimentally derived group index values (see Fig. 1(a)) are used, and the fitted numerically computed group index is used in the case of the model conversion efficiency.

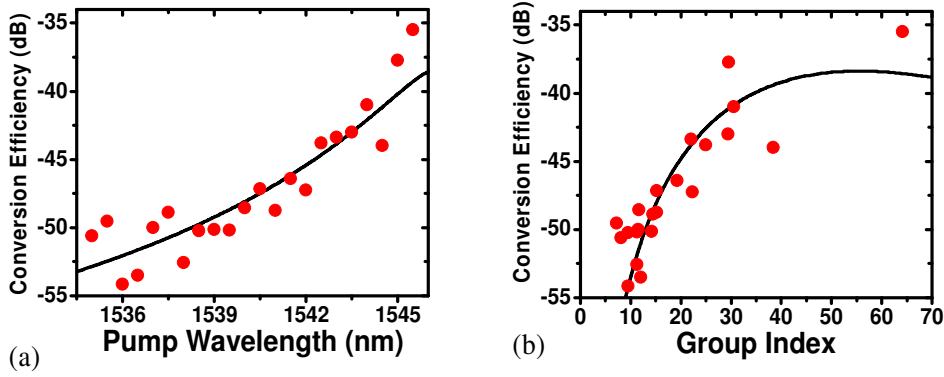

Fig. 4. (a) Four-wave mixing conversion efficiency dependence on pump wavelength (|$\lambda_{signal}$-$\lambda_{pump}$|=0.1nm). (b) Conversion Efficiency dependence on $n_g$ at $\lambda_{pump}$. (points: measurement, solid line: calculation)

The dispersion of the W1 waveguide affords the opportunity to experience very large group indices, however at these wavelengths the corresponding values of $\beta_2$ and $\beta_4$ are very large. It can be seen from Eq. 2 that this will decrease the bandwidth over which FWM can take place. This decrease in bandwidth can be interpreted as the group index increasing the effective length of the waveguide, causing a larger propagated phase mismatch between the interacting wavelengths. This phenomenon can be directly observed in the experimental data in Fig. 3(a) by observing the shrinking signal detuning bandwidth over which idler power is observed as the pump wavelength moves towards $\lambda_{os}$, giving the plot a distinctive triangular appearance. This limitation of the W1 waveguide has been reported previously [33] and can be mitigated by the introduction of regions of low-GVD into the waveguide dispersion by structural tuning [47,48].

To characterize the decrease in the FWM bandwidth as the pump is moved closer to $\lambda_{os}$, the bandwidth of the measured conversion efficiency is plotted in Fig. 5(a). Due to the strong Fabry-Perot oscillations and limited resolution of the experimental data a conventional -3dB bandwidth could not be accurately gauged, so we define the bandwidth as the bandwidth over which the conversion efficiency is within -10 dB of the maximum value for each pump wavelength. This same definition is used to derive the bandwidth from the model. The result shows that the measured bandwidth decreases approximately at a rate of 0.4 nm for every nanometer the pump moves closer to $\lambda_{os}$ and the model matches the experimentally derived values until the last two measured pump wavelengths. This can be attributed to the imperfect match between the measured group index and the shifted numerically derived group index used for the model (illustrated in Fig 1(a)). When the measured bandwidth is plotted with respect to experimentally measured group index, as it is in Fig. 5(b), this discrepancy is removed. The remaining deviations between experiment and model here are attributable to the finite signal wavelength step (0.1 nm) and the Fabry-Perot induced uncertainty in the group index when using the phase shift method [49].

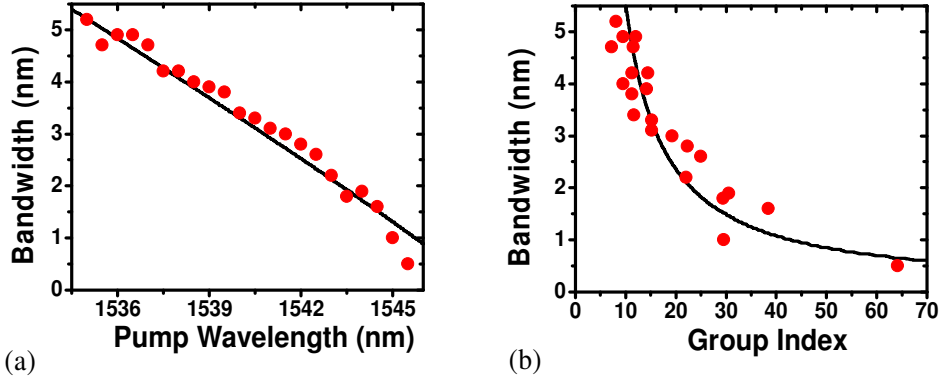

Fig. 5. The -10-dB Bandwidth (defined in the text) dependence on (a) $\lambda_{pump}$ and (b) $n_g$ (at $\lambda_{pump}$).

As was noted previously, the simple model employed in this work uses a linear $n_g$ scaling of the linear propagation loss: $\alpha^* = \alpha_0 \left( n_g / n_0 \right)$, where $\alpha_0$ was found to be 10 dB/cm. This choice of scaling was made based on the fact that it best fit the data measured, as can be seen in Fig. 6 where the data presented in Fig. 4(b) is shown with the conversion efficiency of the model for different propagation loss $n_g$ scaling factors with all other factors being equal. A number of recent experiments and theoretical investigations on the exact scaling of loss with respect to group index in photonic crystal waveguides has provided a number of differing opinions on the subject [41,51,50,52], with several reports on the fabrication disorder-induced loss in photonic crystal waveguides, like that of regular channel waveguides [53], scaling linearly with $n_g$. In addition to this out of plane scattering, there is a scattering from the forward propagating mode into the backward propagating mode which scales with $n_g^2$. However, fabrication disorder can cause coherent scattering within and between unit cells of the waveguide along the direction of propagation [43]. In addition, there is a modification of the Bloch mode shape with respect to frequency which as the effect of increasing the disorder sampling. These two factors have been used to suggest that the Beer-Lambert law and the simplistic view of relating group index scaling of loss fail for large group indices [54]. In regards to this experiment, since the majority of the data collected was in the $n_g$ less than 40 regime, the dominant scaling of propagation loss observed was linear.

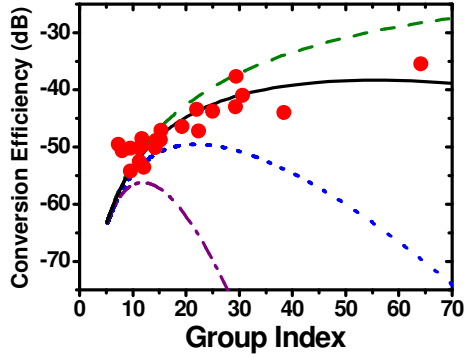

Fig. 6. Conversion Efficiency dependence on $n_g$ at $\lambda_{pump}$ for different propagation loss $n_g$ scaling factors: $\alpha^* = \alpha_0 \left( n_g / n_0 \right)^\eta$. (dash) $\eta = \sqrt{2}/2$ (solid) $\eta = 1$ (dot) $\eta = \sqrt{2}$ (dash-dot) $\eta = 2$. (points: measurement)

One possible source of experimental uncertainty with regard to the group index at $\lambda_{pump}$ comes from the thermal shift of $\lambda_{os}$ due to the presence of the pump beam. The thermal heating of the sample caused by the pump beam would cause a red-shift in the mode onset wavelength in contrast to the data presented in Fig. 1(a) which was collected at low optical powers. The waveguide is more susceptible to this effect closer to $\lambda_{os}$ due to the large rate of $n_g$ increase and the large oscillations of transmission due to the Fabry-Perot resonances at these wavelengths. However, due to the agreement between the simulated measured bandwidth (see Fig. 5) which is directly dependant on the value of $\beta_2$ [55], we are confident that thermal shift of the group index is minimal.

## 5. Conclusion

We have experimentally investigated the group index enhancement of four-wave mixing in a W1 silicon photonic crystal waveguides. A 0.44 mm long waveguide exhibited a maximum conversion efficiency of -36 dB using a coupled pump power of 14 dBm. Over the wavelengths examined, the conversion efficiency was measured to increase by over 12 dB due to a group index enhancement of the third-order nonlinearity. A corresponding decrease in the conversion efficiency bandwidth, from 5 nm to 0.5 nm, was also observed. Both of these experimental observations match well with a simple numerical model of four-wave mixing in photonic crystal waveguides which accounts for both nonlinear and propagation loss scaling with group index. The results presented here reinforce the slow-light nonlinear enhancement possible within silicon photonic crystal waveguides.

**Acknowledgments**

The authors acknowledge discussions with C. A. Husko, M.-C. Wu, and N. C. Panoiu. This work is supported by NSF (ECCS-0622069 and ECCS-0747787).